\documentclass[a4paper]{article}

\usepackage{icrc2013}

\usepackage[english]{babel}

\title{Towards SiPM camera for current and future generations of Cherenkov telescopes}

\shorttitle{Tests for SiPMs for MAGIC and CTA}

\authors{
Daniel Mazin$^{1}$,
Priyadarshini Bangale$^{1}$,
Julian Sitarek$^{2}$,
Juan Cortina$^{2}$,
David Fink$^{1}$,
J\"{u}rgen Hose$^{1}$,
Jose Maria Illa$^{2}$,
Eckart Lorenz$^{1}$,
Manel Mart\'{\i}nez$^{2}$,
Uta Menzel$^{1}$,
Razmik Mirzoyan$^{1}$, and
Masahiro Teshima$^{1}$
}

\afiliations{
$^1$ Max-Planck-Institut f\"ur Physik, D-80805 M\"unchen, Germany \\
$^2$ Institut de F\'{\i}sica d'Altes Energies (IFAE), E-08193 Bellaterra, Spain \\
}

\email{mazin@mpp.mpg.de}

\abstract{
So far the current ground-based Imaging Atmospheric Cherenkov Telescopes
(IACTs) have energy thresholds in the best case in the range of ~30 to 50 GeV
(H.E.S.S. II and MAGIC telescopes).  Lowest energy gamma-ray showers produce
low light intensity images and cannot be efficiently separated from dominating
images from hadronic background.  A cost effective way of improving the
telescope performance at lower energies is to use novel photosensors with
superior photon detection efficiency (PDE).  Currently the best commercially available superbialkali
photomultipliers (PMTs) have a PDE of about 30-33\%, whereas the silicon
photomultipliers (SiPMs, also known as MPPC, GAPD) from some manufacturers show
a photon detection efficiency of about 40-45\%.  Using these devices can lower
the energy threshold of the instrument and may improve the background rejection
due to intrinsic properties of SiPMs such as a superb single photoelectron
resolution.  Compared to PMTs, SiPMs are more compact, fast in response,
operate at low voltage, and are insensitive to magnetic fields.  SiPMs can be
operated at high background illumination, which would allow to operate the IACT
also during partial moonlight, dusk and dawn, hence increasing the instrument
duty cycle.  We are testing the SiPMs for Cherenkov telescopes such as MAGIC
and CTA.  Here we present an overview of our setup and first measurements,
which we perform in two independent laboratories, in Munich, Germany and in
Barcelona, Spain.  }

\keywords{MAGIC, CTA, Cherenkov telescopes, photodetectors, SiPMs, avalanche photodiodes, cross-talk, photodetection efficiency}

\begin{document}
\maketitle

\section{Introduction}
Many of the current astro-particle physics experiments exploit detection
technique based on detection of optical or UV light flashes produced by high
energy particles in various processes.  For example, ground-based telescopes can
detect fluorescence (e.g.\ Pierre Auger Observatory, \cite{auger}) or Cherenkov
(e.g.\ the MAGIC telescopes, \cite{upgrade}) light produced by particles in 
atmospheric air showers, neutrino detectors such as IceCube \cite{icecube}
detect Cherenkov light flashes produced by muons in ice or water.  Also some space-based
instruments (e.g.\ calorimeter in Fermi-LAT \cite{fermi}) detect high energy
particles and gamma rays by measuring particle showers produced in the calorimeter.

The energy threshold of the Imaging Atmospheric Cherenkov Telescopes (IACTs) is
determined by the amount of Cherenkov photons that a telescope can detect above
the level of the fluctuations of the night sky background (NSB) light.  It is
essential to achieve low energy threshold in order to be able to observe and
study some classes of objects, such as pulsars which have intrinsically very
soft spectra, or distant active galactic nuclei, in which the higher energy
part of the emission is absorbed by low energy photon fields on the way to the observer. 

So far the current ground-based IACTs have the energy threshold in the range
of ~30 to 50 GeV (H.E.S.S. II and the MAGIC telescopes).  The future CTA
project \cite{cta} is aiming at obtaining a trigger threshold of few tens of
GeV.  Moreover, the images produced by gamma rays with energies close to the
threshold are badly reconstructed, among others because of scarce 
light available.  This results in very poor hadron background suppression at
those energies.  The noise produced by NSB in a single camera pixel is only
weakly dependent on the mirror area of the telescope because the flux of NSB
photons is isotropic and constant.  On the other hand the amount of detected
photons from extended air showers increases linearly with the area of the
mirror dish.  Moreover, for arrays of IACTs increasing the amount of light
detected by individual telescopes will result in more telescopes detecting a
given shower.  Having several images of the same air shower allows one to
further improve the background suppression and reconstruction of the arrival
direction and energy of the incident gamma ray.  Thus going for larger mirror
dishes would further decrease the energy threshold and improve background
reduction.  However, the construction of such large telescopes becomes too
expensive and technically challenging. 

Another way of improving the sensitivity at lower energies is to use novel
photo-sensors with superior quantum efficiency (QE) and superior timing
resolution.  One of a promising possibility are the silicon photomultipliers
(SiPMs, \cite{Bondarenko, Buzhan, Golovin, Otte}).  In fact, SiPMs are already
used in a small Cherenkov Telescope, FACT \cite{fact}.  In comparison with
photomultipliers (PMTs) used in most of the current IACT experiments SiPMs are comparably fast,
operate at low voltage (a few tens of V), and insensitive to magnetic fields.
In general, they have good photon detection efficiency (PDE), which eventually
in future could lower the energy threshold of the IACTs, and they can be
operated at high background illumination, hence can increase the instrument
duty cycle.  So far, the conventional PMTs have a QE of about 30-33\%
\cite{razmik} but newer PMTs show peak QE of about 40--45\% (R. Mirzoyan, private communication). 
Theoretically, SiPM
can beat even this, reaching PDE of 60-70\%. With a higher QE of the
photo sensors, one can lower the energy threshold of the instrument and improve
the background rejection.

We are working on a development of modular SiPM clusters with a cluster size to cover a detector area of about 50\,cm$^2$. 
In the first step, which is the focus of this paper, we evaluate some selected SiPM available on the market and compare their
response with the one from PMTs. We evaluate breakdown voltage, cross-talk, gain linearity, and PDE
as a function of light wavelength. All measurements are performed at room temperature of 26$^{\circ}$\,C unless stated otherwise.

\section{SiPM basics}\label{basics}

\begin{figure}[t]
  \centering
  \includegraphics[width=0.5\textwidth]{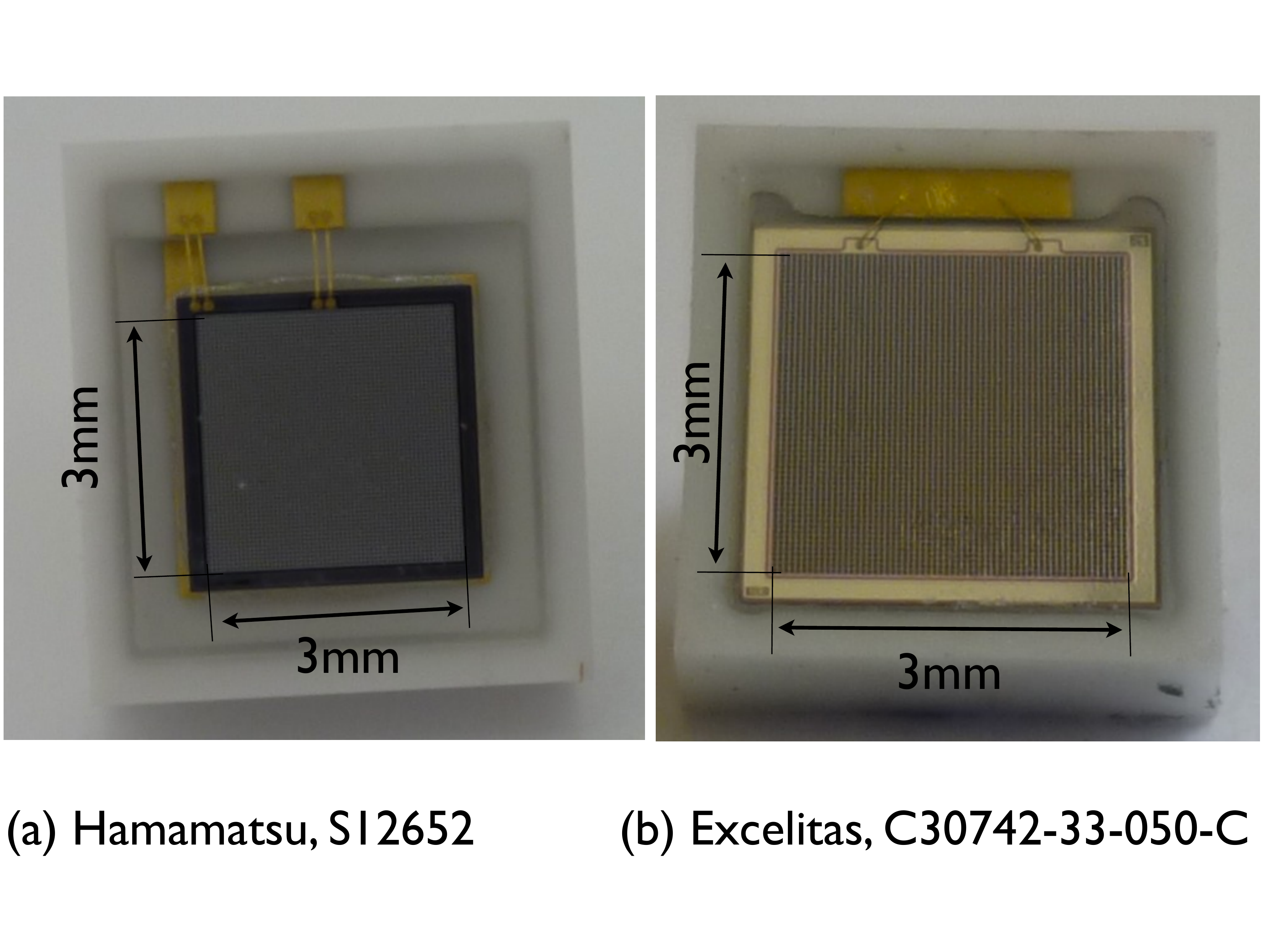}
  \caption{Two of the four SiPM devices used in the measurements: (a) Hamamatsu 3$\times$3 mm (S12652-A0013), 
(b) Excelitas 3$\times$3 mm (C30742-33-050-C, A0896). Cell size is 50\,$\mu$m for all of the tested SiPMs here.}
  \label{sipm}
 \end{figure}

A SiPM is a matrix of cells, each being an avalanche photodiode (APD),
connected to a common anode bus.  The cells are operated in the limited Geiger
mode, i.e.\ biased a few Volts above the breakdown voltage.  A single photon can
release a Geiger avalanche process and produce a well defined signal from a single cell.  The
signals from all cells are added up on the bus and thus the output signal of a
SiPM is the summed up signal of all coinciding in time ``fired'' cells.  The
individual cells by themselves behave as a binary device (i.e.\ signal generated
by two consecutive photon will be the same as of a single one), but due to
large number of them (of the order of 1000 cells per SiPM) the whole device
behaves in analog way. 

The single photon response is partially affected by the effects of the optical
cross-talk and afterpulses.  During the avalanche in a cell of SiPM some of the
carriers can be trapped for some time in the impurities of the semiconductor
crystal structure. If such carrier is released after the cell recharges it will
generate a new avalanche, with a signal of 1 photon or a fraction of it (in
case the cell did not had time to recharge fully after the original discharge).
After pulses depends on temperature, pixel recovery time (more contribution for
shorter pixel recovery time) and overvoltage. 

Moreover, during the discharge of a cell light is emitted. 
Some of it may travel to a neighboring cell and trigger another Geiger avalanche there. 
Such an effect (called optical crosstalk) depends on the gain of the device, and thus also on the temperature.
The crosstalk will clearly modify the Poissonian statistics of the number of registered photons.
The probability $P(n)$ to observe in total $n$ fired cells is given by (see \cite{gapd}):

\begin{equation}
 P(n) = \sum\limits_{j=1}^{n} P^{0}_{j}(1-\varepsilon)^{j} \varepsilon^{n-j} \left(\!\!
    \begin{array}{c}
      n-1 \\
      j-1
    \end{array}
  \!\!\right),
\end{equation}
%
where $P^{0}_{j}$ is a Poissonian probability to fire $j$ cells in the original
light pulse (before the cross-talk) and $\varepsilon$ is the crosstalk
probability. \\ \\
The PDE of a SiPM is a function of several factors: ratio of the single cell
light sensitive area to its total area (so-called fill factor), probability for starting a Geiger
avalanche (which depend on overvoltage and temperature), transport of impinging
photons into the sensitive volume, and an intrinsic quantum efficiency \cite{gapd}.
The last 3 factors depend on the wavelength of the photon.

%
%


\section{Measurements and analysis}\label{analysis}
\begin{figure}[t]
  \centering
  \includegraphics[width=0.45\textwidth]{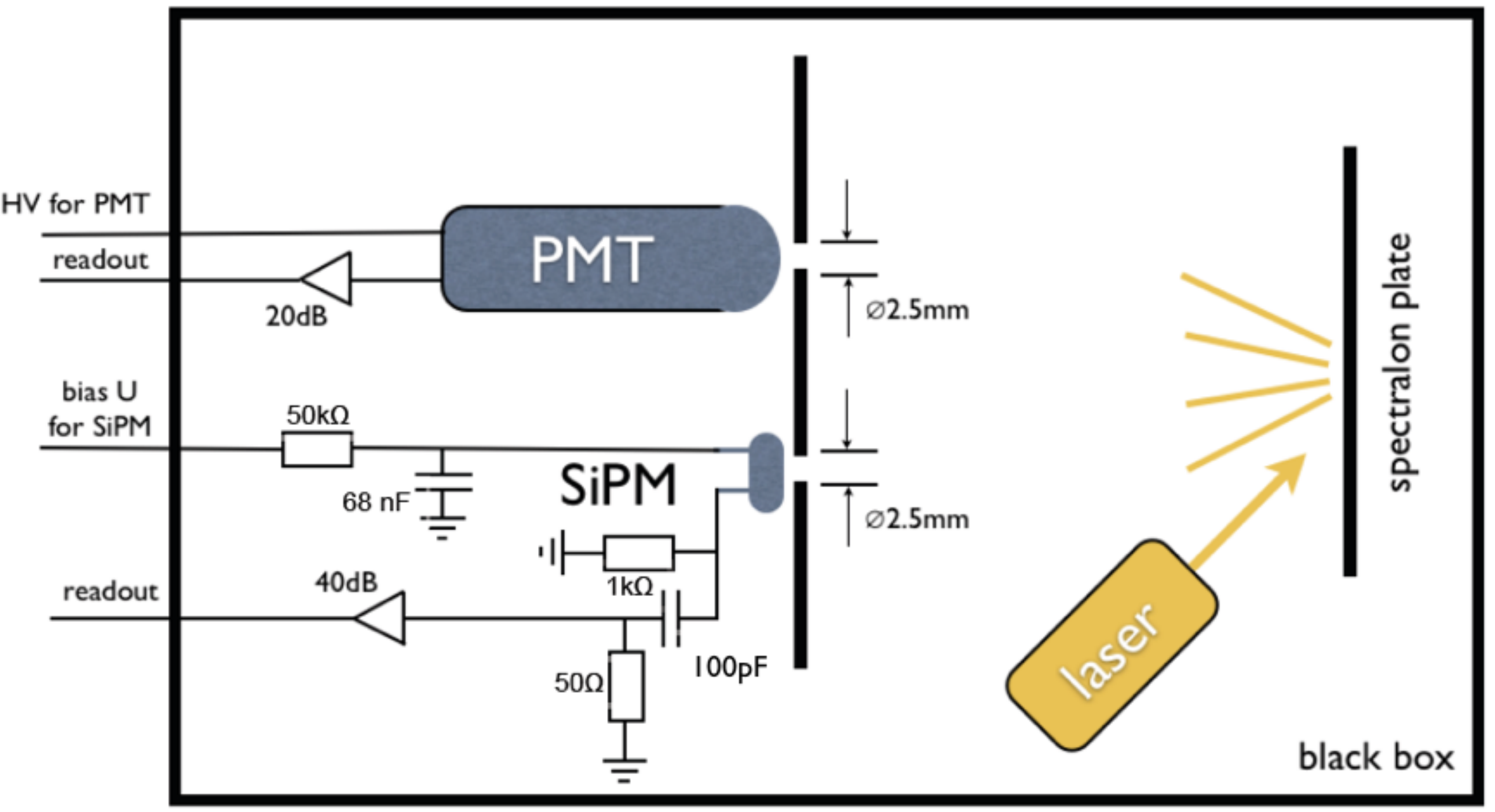}
  \caption{Sketch of the measurement setup.}
  \label{setup}
 \end{figure}

Here we have studied performance of four SiPMs, which were developed by two
companies: Excelitas\footnote{http://www.excelitas.com} and 
Hamamatsu\footnote{www.hamamatsu.com}. Those are:
\begin{itemize}
 \item Hamamatsu, 3mm x 3mm, Nr 14, Low After Pulsing (LAP) unit through high purity silicon, no trenches;
 \item Hamamatsu, 3mm x 3mm, S12652 - A0013, trenches introduced to suppress cross-talk;
 \item Excelitas, 3mm x 3mm, C30742-33-050-C A0896, trenches introduced to suppress cross-talk;
 \item Excelitas, 6mm x 6mm, C30742-66-050-C, E1479, trenches introduced to suppress cross-talk;
\end{itemize}
All measured SiPMs have a cell size of 50\,$\mu$m. Hamamatsu-S12652 and Excelitas-C30742-33-050-C
are shown in Figure \ref{sipm}. The SiPMs are in ceramic packaging and contacts for bias voltage and
signal are visible on the top part.

Figure \ref{setup} shows the measurement setup for comparing properties of
SiPM with a calibrated PMT. We used 1.5\,inch R11920-100 calibrated PMT 
from Hamamatsu as a reference sensor. The setup is placed in a light-tight box.
The light from a laser (FWHM of about 50\,ps) operating in a pulse-mode with a pulsed diode
driver PDL800-B from Picoquant was delivered to an optical filter. Then the
light was routed to the light-tight box, where it illuminates a spectralon plate, which is used as a diffuser.
The SiPM and the PMT are mounted behind a plate at the same distance (about 50\,cm) from the spectralon plate.
Both photosensors are equipped with a 2.5\,mm diameter diaphragm to ensure the entrance windows
(i.e.\ the amount of photons) for the two devices are identical. There is also a possibility to swap the PMT and the SiPM
in order to calibrate the non-uniformity of the diffuser or/and imperfection of the mechanical alignment.
The signal of the SiPM is filtered by a high pass filter to shorten the signal and then amplified by 
a Femto amplifier\footnote{http://www.femto.de/} with a gain of 40dB. 
The PMT signal is amplified by a PACTA pre-amplifier\footnote{The PACTA pre-amplifier, http://iopscience.iop.org/1748-0221/7/01/C01100}
with a again of 20dB. Both amplified signals are terminated with 50$\Omega$ resistors 
and sampled with a Tektronix DPO 7254C oscilloscope. 
Detection of light at a few wavelengths in the range of 256-598 nm were investigated. When
possible, several light intensities were applied for each wavelength.
For each data run we save complete waveforms of pulses for 86000 events with sampling
frequency of 10GS/s. Using the data, we determine FWHM of the pulses,
the mean number of photoelectrons (phe) for a given light intensity, conversion factor from extracted charge into phe's,
and the cross-talk probability.

\begin{figure}[t]
  \centering
  \includegraphics[width=0.5\textwidth]{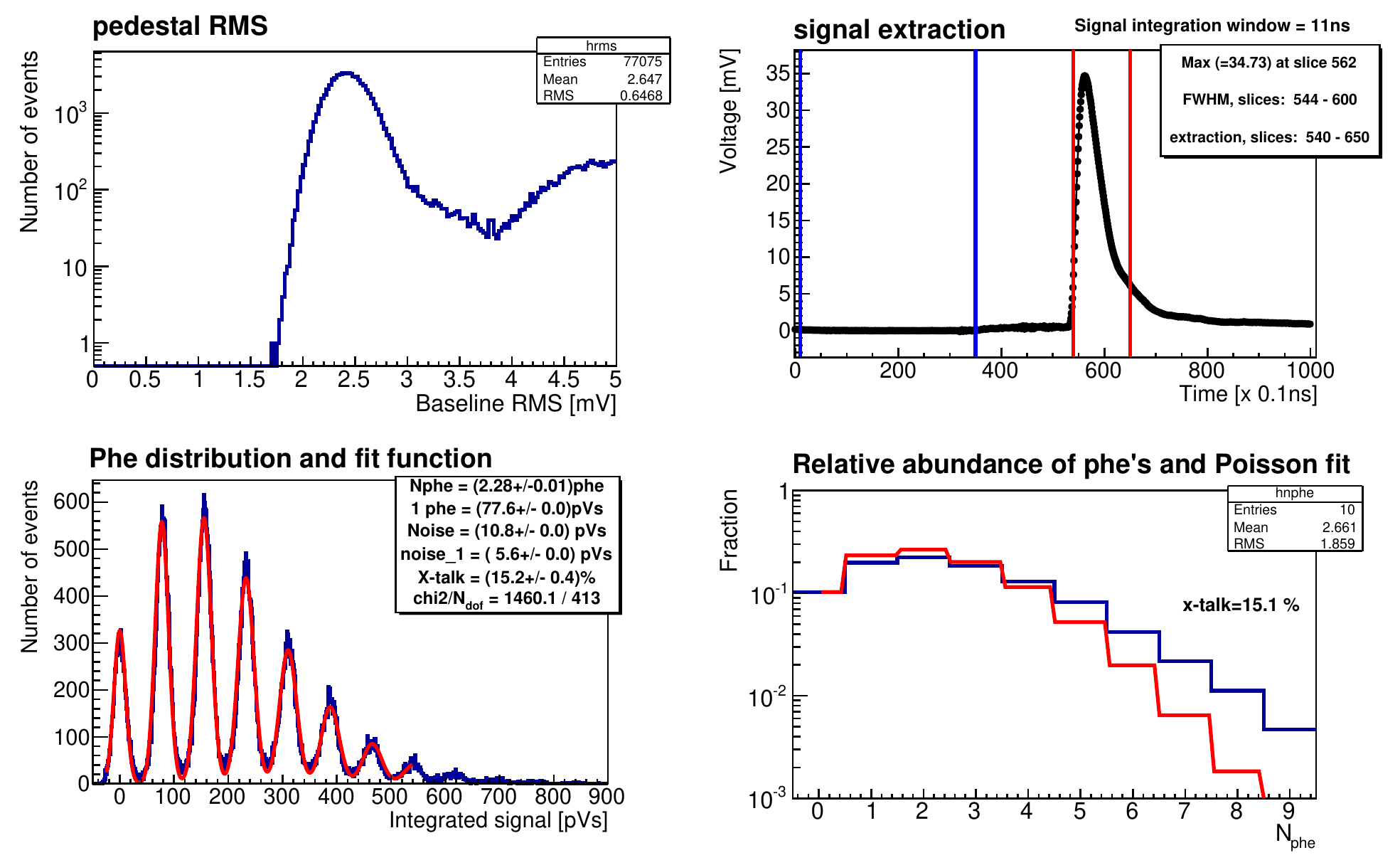}
  \caption{Example of measurements. 
{\it Top left:}
 The pulse shape of Hamamatsu 3$\times$3 mm LAP (Nr 14) SiPM,
which is signal averaged over all waveforms. First
34 ns (blue lines) were used to determine the baseline.
{\it Top right:} Baseline RMS estimated from beginning of the readout window.
{\it Bottom left:}
A single photoelectron spectrum measured at low intensity light from the laser. Data in blue,
fit in red.
{\it Bottom right:}
Probability distribution to measure $n$ phe's. Data in blue, pure Poissonian fit (no cross-talk) 
to the zero's and the total number of events is in red.}
  \label{measure}
 \end{figure}

In Figure~\ref{measure} we show some histograms from the analysis chain on an example of Excelitas 
C30742-66-050-C, E1479 operated at a bias voltage of V=101.5V and using a laser light of 405nm.
Baseline RMS showed in the top left panel has a main Gaussian
component and a tail caused by dark rate phe's occurring in
the baseline estimation region. On the top right we show a mean pulse shape from 86000 measurements.
First 34 ns (region between the vertical blue lines) were used to determine the baseline.
The signal is then integrated (after baseline subtraction) in the region between the vertical red lines (11\,ns).
Different tested sensors showed verying values of full width half maximum (FWHM) of the light pulses, so we
optimized size of the extraction window for all of them separately. The width and the position of the integration window is kept 
fixed for a given SiPM and a given wavelength.

We fill a histogram with such integrated signals (blue histogram in the bottom left plot). 
The distribution is then fitted with a single function being a sum of
equally spaced Gaussians. The RMS of the individual Gaussians are
$\sigma_i = \sqrt{\sigma_0^2 + i * \sigma_r^2}$, where $\sigma_0$ is the
pedestal noise RMS (estimated from dedicated runs without light
pulses), and $\sigma_r$ is an additional noise component coming from differences between individual cells in SiPM
device, and $i$ is the number of generated phe.
The integrals of individual Gaussians are computed from Poissonian
distribution with a correction for the optical cross-talk factor
according to \cite{gapd}. From the fit value we obtain the average, primary (i.e. without the
phe generated in the cross-talk process) number of phe's, conversion factor (distance between the individual peaks), 
the $\sigma_r$ parameter, and the cross-talk probability.
The resulting fit parameters are displayed in the inlay. 

As an additional check of the robustness of the method we also tried
to split the fitting procedure into two steps.
In the first step we fit a function being a sum of Gaussians as
described above, but allowing the normalizations of individual
Gaussian peaks to be arbitrary. In the second step we fit the normalizations obtained from the first step with a Poissonian
distribution with the optical cross-talk correction. Both analysis methods give very similar results.

The number of events in the 
phe distribution is then integrated for each peak and plotted (normalized) in the bottom 
right (blue curve). A Poissonian fit is performed to the zero's peak (red line) while keeping the total number of events fixed,
resulting in an estimation of the cross-talk, which is defined as the missing number of events with 1 phe.
This is an alternative estimation of the cross-talk compared to the one done in a global fit
(bottom left distribution). Both methods gave consistent results for all cases where fits converged well.

\section{First results and discussion}

\begin{figure}[t]
  \centering
  \includegraphics[width=0.5\textwidth]{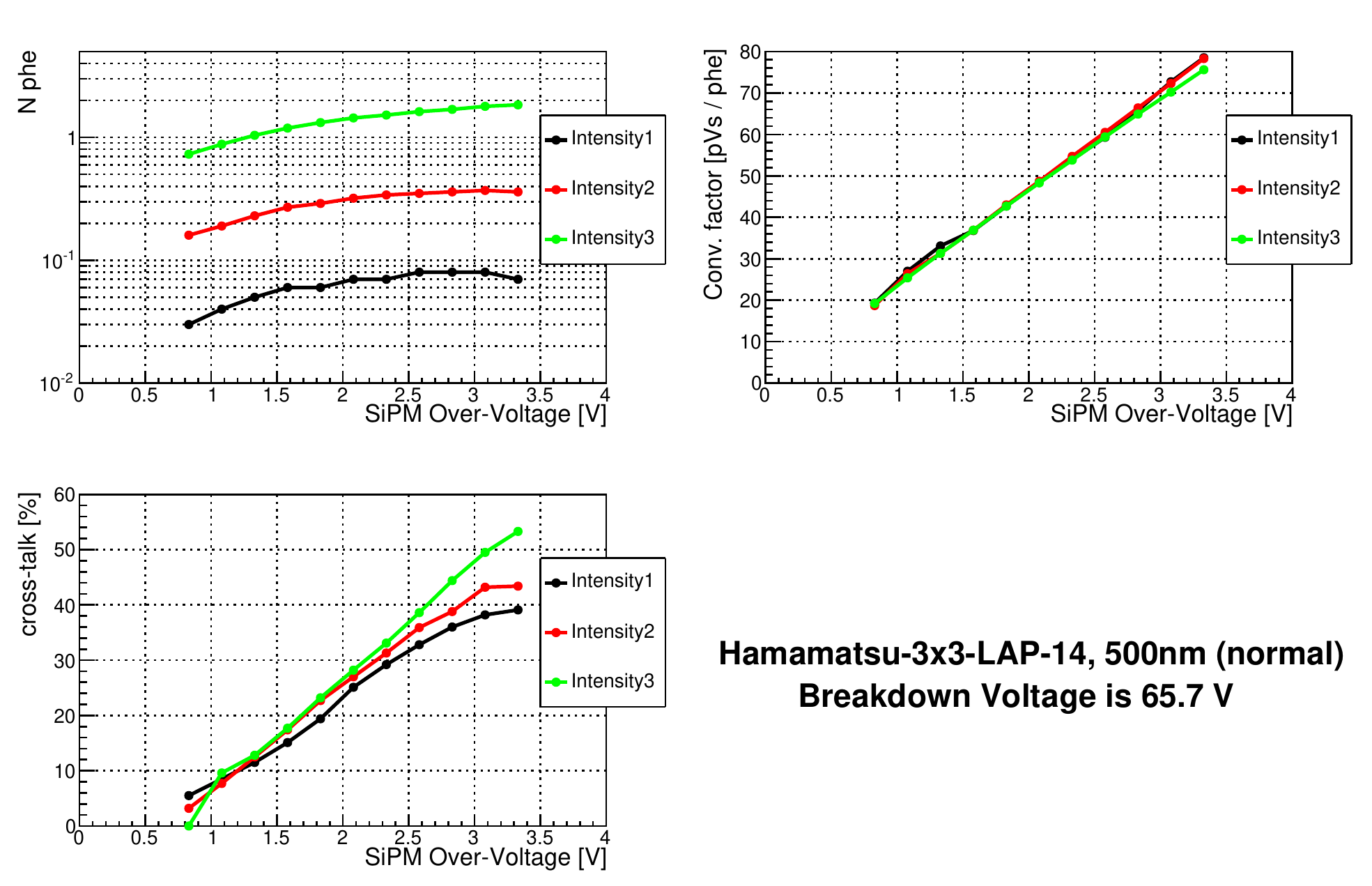}
  \caption{Series of measurements of Hamamatsu 3\,mm$\times$3\,mm LAP (Nr 14) taken for 
3 different light intensities ($\lambda=$500nm) and a span of applied bias voltages.}
  \label{results}
 \end{figure}

In Figure \ref{results}, we show results of a series of measurements of Hamamatsu 3$\times$3 mm LAP (Nr 14) taken
at 500nm for 3 different light intensities and a span of applied bias voltages.
The measurements were performed at the temperature of 28$^{\circ}$\,C. The number of phe's is slowly raising with the applied voltage
due to increasing probability to produce a Geiger avalanche. From extrapolation of the conversion factor (i.e. integrated signal
corresponding to one phe) vs overvoltage curve to conversion factor $=$ 0 we find that, the breakdown voltage for this device is 65.7 V.
As expected the cross-talk probability raised with the applied overvoltage. From the measurements performed at different light intensities, we
conclude that the relative systematic error in the estimation of the cross-talk due to simplifications in the fitting function is about 10\%.

\begin{figure}[t]
  \centering
  \includegraphics[width=0.5\textwidth]{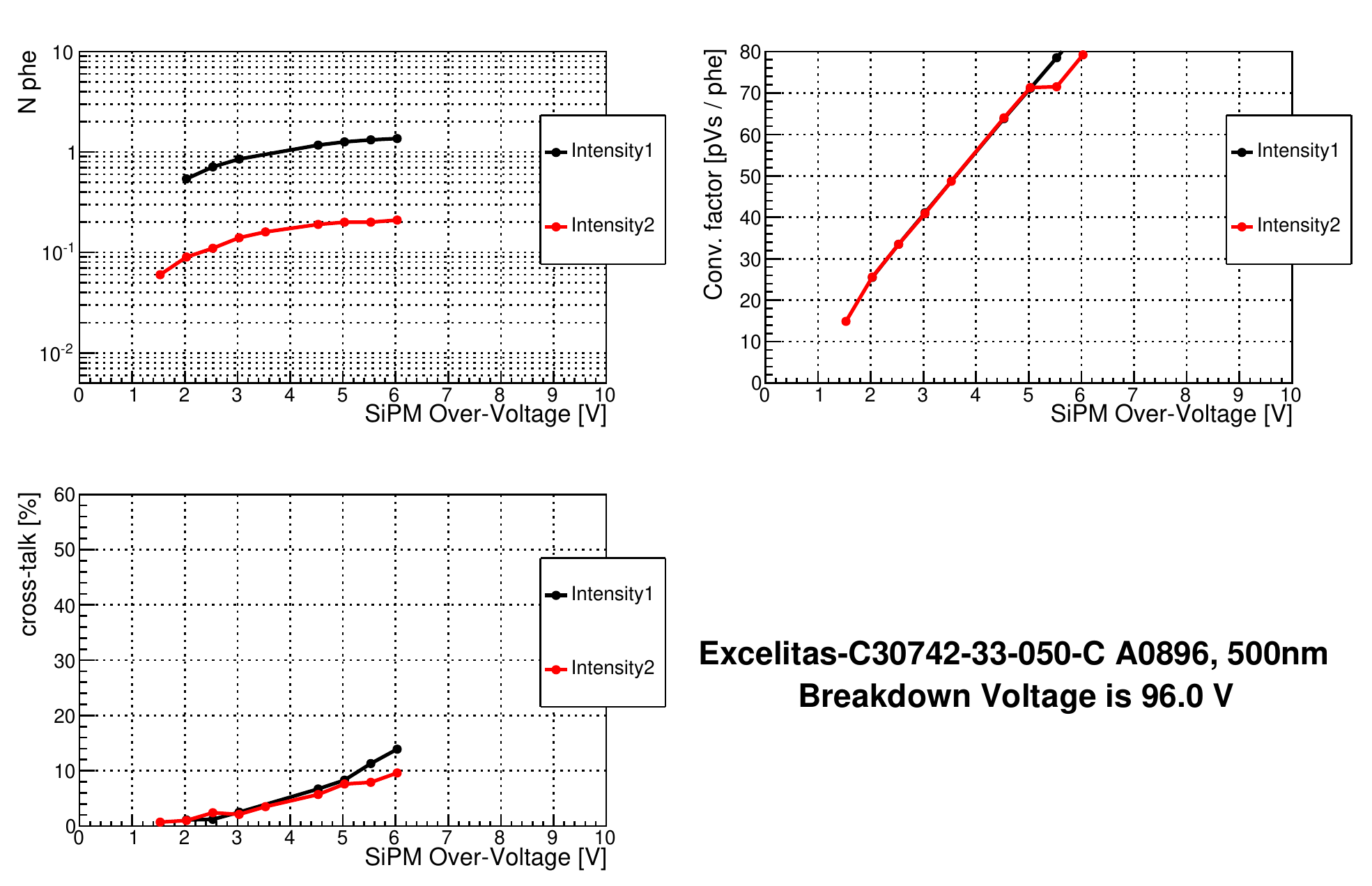}
  \caption{Series of measurements of Excelitas-C30742-33-050-C A0896 3\,mm$\times$3\,mm taken for 
two different light intensities ($\lambda=$500nm) and a span of applied bias voltages.}
  \label{resultsEx}
 \end{figure}

In Figure \ref{resultsEx}, we show results of a series of measurements of Excelitas C30742-33-050-C A0896
for 500nm two different light intensities and a span of applied bias voltages.
In comparison to the Hamamatsu SiPM (above), the break down voltage of the Excelitas device is higher (96.0 V)
and it is possible to operate it at slightly higher percentage overvoltage: Excelitas can operate at 6V (6.25\%)
whereas the Hamamatsu SiPM can operate at 3.25V overvoltage (4.9\%). Both devices seem to almost reach the plateau
in the PDE curve. The other difference is clearly the cross-talk probability: whereas the Excelitas SiPM is below 10\%
crosstalk for overvoltage of 5\%, Hamamatsu is reaching 40-50\% crosstalk at the same overvoltage.
This is clearly a result of using trenches to suppress the cross-talk.
We see, however, that in the second Hamamatsu device (S12652 - A0013), where trenches also have 
been used, the cross-talk probability is similar to the one of Excelitas (taken at fixed percentage of overvoltage).

In this first study we see that the new generation of SiPM from Hamamatsu and Excelitas
have succeeded on two areas: 1) the cross-talk is significantly lowered (by using trenches)
by a factor 2--3 and 2) the operation range is increased such that the SiPM can be operated at 5-6\% relative overvoltage
(see \cite{mppcstudy} for comparison),
which provides a relatively high PDE. Qualitatively we see that the PDE of SiPM is better that the one of the  
R11920-100 PMT at wavelengths above 350nm and still worse at lower wavelengths.
Detailed results on PDE, cross-talk and temperature dependence are a subject of
a dedicated publication.
Another major result is that the pulse shapes of the SiPMs of these study are sufficiently short,
with rise time below 2\,ns and FWHM of 7-15\,ns (after the differentiating of the signal). 

The current generation of SiPMs have still a few disadvantages. Such as,
afterpulsing, cross talk, only small sizes of the devices are available (most
of the devices have sizes of a few mm) and also the photon detection efficiency
(PDE) is limited by the geometrical factor (area of the active surface of the
device to the total one). The SiPMs performance (gain, dark rate)
depends strongly on the bias voltage and the temperature.  This requires either
extremely stable working conditions, which is often not feasible, or feedback
circuits. Another possibility is to operate SiPM in the saturation regime,
where at least PDE is insensitive to temperature changes. Finally, the spectral response of SiPMs is not yet optimized
for the Cherenkov spectrum at usual altitudes (around 2000\,m above the sea level):
the SiPMs have rather low PDE in UV regime (this must be improved) and high sensitivity
in the infrared, where the signal from the NSB dominates. These two facts
currently limit the signal to noise ratio of the SiPM and will be addressed in the next generation of SiPMs.

\vspace*{0.5cm}
\footnotesize{{\bf Acknowledgment:}{The authors would like to warmly thank for excellent technical support at
 Max-Planck-Institut f\"ur Physik, Munich and at IFAE, Barcelona. 
A very good cooperation with companies Hamamatsu and Excelitas is gratefully acknowledged. }}

\end{document}